# Machine-learning-identified two-dimensional van der Waals multiferroics for four-state nonvolatile memory

Zhibin Tan,[1] Tao Wang,[2] and Hao Jin[1,*]

1. College of Physics and Optoelectronic Engineering, Shenzhen University, Shenzhen, China.

2. International School of Microelectronics, Dongguan University of Technology, Dongguan, China.

E-mail: jh@szu.edu.cn

**Abstract**

Two-dimensional (2D) van der Waals (vdW) multiferroics offer an attractive platform for four-state nonvolatile memory by combining switchable ferroelectric polarization and magnetization within a single material system. However, their development is hindered by the scarcity of synthesizable candidates and the lack of non-destructive readout schemes. Here, we combine machine-learning screening with first-principles calculations to explore the 2D vdW $ABC_2X_6$ family and identify a set of high-confidence multiferroic candidates. Among them, $AuCrP_2S_6$ monolayer emerges as a representative system with a ferromagnetic ground state, a sizable out-of-plane polarization of 7.46 pC/m, and a moderate ferroelectric switching barrier of ~130 meV/f.u. Moreover, the nonlinear optical response mediated by the bulk photovoltaic effect (BPVE) in $AuCrP_2S_6$ provides a dual-channel probe of the ferroic orders, in which the polarization direction governs the photocurrent sign while the magnetic order selects the spin channel via robust exchange splitting. This intrinsic coupling enables the non-destructive readout of four logic states within a single atomic layer, thereby providing a practical blueprint for next-generation multistate optoelectronic memory.

The continued pursuit of high-density nonvolatile memory in the post-Moore era has driven material exploration toward the two-dimensional (2D) limit.[1-4] Among the vast library of van der Waals (vdW) materials, 2D multiferroics that simultaneously host ferroelectric (FE) and magnetic orders are emerging as a promising platform for next-generation spintronics.[5-8] By integrating FE polarization ($P$) and magnetization ($M$) in a single atomic layer, these materials promise low-power multistate memory encoded by distinct ($P$, $M$) configurations.[9,10] However, despite the growing families of 2D ferromagnets and ferroelectrics, intrinsically multiferroic monolayers are exceptionally rare.[11-13] This scarcity stems from a fundamental electronic incompatibility, where ferroelectricity and magnetism often impose competing electronic requirements.[14-17] Consequently, identifying intrinsically multiferroic 2D platforms that simultaneously stabilize FE and magnetic orders remains a central challenge.

In recent years, data-driven screening has become a powerful tool for navigating vast compositional spaces, yet predicting synthesizability is a nontrivial task.[18-20] A fundamental reason is the inherent asymmetric data landscape: experimentally synthesized materials provide reliable positive examples, while the vast majority of hypothetical structures are merely unlabeled rather than definitive negatives for synthesizability. This positive-unlabeled (PU) data bias undermines traditional supervised learning approaches that rely on both positive and negative samples.[21,22] The issue is further compounded for 2D materials, where experimentally validated datasets are sparse. Even when synthesizable candidates are identified, a pivotal bottleneck for practical multiferroic memory lies in the readout mechanism. Conventional electrical detection of FE states is often destructive and therefore requires a subsequent re-writing cycle.[23,24] Moreover, distinguishing coupled ($P$, $M$) states is frequently compromised by signal crosstalk.[16,25-27] Accordingly, it is highly desirable to identify 2D multiferroic platforms that combine experimentally plausible candidate materials with intrinsic non-destructive readout functionalities. To this end, we combine machine-learning screening with first-principles calculations to explore the chemically versatile $ABC_2X_6$ vdW family (where A/B denote distinct cations, C is a pnictogen, and X is a chalcogen), which ranks candidates by likelihood and yields a high-confidence pool for subsequent screening. Guided by this framework, monolayer $AuCrP_2S_6$ is identified as a promising 2D multiferroic semiconductor with a ferromagnetic ground state, a large out-of-plane polarization, and a moderate switching barrier. Building on this candidate, we further propose an optoelectronic readout scheme based on the nonlinear optical response via bulk photovoltaic effect (BPVE). Our results demonstrate that the photocurrent serves as a sensitive dual-channel probe, whose signal polarity and spin character are uniquely locked to the ferroic orders. This mechanism enables a deterministic, non-destructive decoding of four logic states ($\pm P$, $\pm M$), highlighting the potential of 2D multiferroics for multistate nonvolatile memory.

To navigate the vast $ABC_2X_6$ chemical space under the constraint of PU data bias, we employ a semi-supervised PU-bagging strategy based on a crystal graph convolutional neural network (CGCNNs), as illustrated in **Figure 1a**. Instead of treating unlabeled structures as true negatives, each sub-classifier is trained with iteratively resampled pseudo-negative subsets, and the final confidence-level (CL) score is obtained by ensemble aggregation.[28,29] As illustrated in **Figure S1** of the supplementary material, the model achieves convergence with 100 sub-classifiers, enabling robust ranking while reducing sensitivity to any single pseudo-negative draw.

Given the scarcity of experimental 2D data, we further introduce transfer learning to bridge the structural knowledge from 3D bulk crystals to 2D monolayers.[31-33] We first pre-train the CGCNNs on large-scale bulk entries from the Materials Project,[34] where synthesizability is supported via cross-referencing with the Inorganic Crystal Structure Database (ICSD),[35] and then fine-tune the model on the target 2D $ABC_2X_6$ dataset using parameter regularization (see **Figure 1b**). This two-stage strategy allows the network to inherit general chemical/structural features while adapting to the 2D vdW layered geometry. CGCNNs are adopted here because they extract structure-sensitive representations directly from the crystal graph without handcrafted descriptors.[36-38] The learned embeddings capture local coordination and polyhedral distortions (e.g. $[C_2X_6]^{4-}$ octahedral deformation) as well as the spatial matching of cations, thereby enabling efficient down-selection from thousands of enumerated candidates to a high-confidence pool (see **Figure 1c**). Detailed model architecture is provided in **Section S1** of the supplementary material.

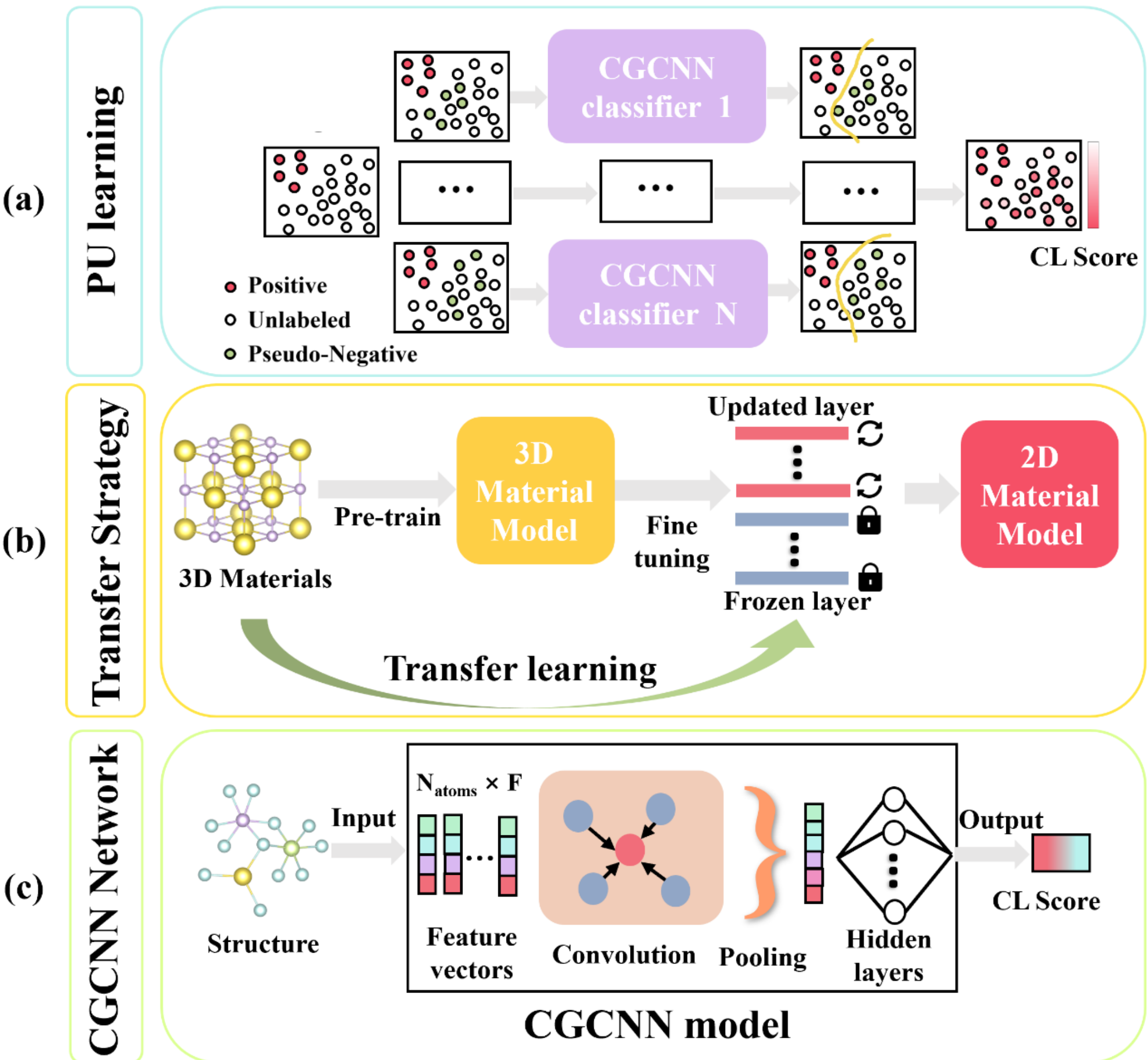

**Figure 1.** (a) Illustration of the PU-learning ensemble, where unlabeled data is iteratively sampled as pseudo-negatives to train multiple classifiers, aggregating their predictions into a final CL score. (b) Transfer learning strategy bridging the knowledge gap between 3D bulk crystals and 2D monolayers by pre-training on bulk data and fine-tuning on the 2D dataset with parameter regularization. (c) The CGCNNs architecture that maps the input crystal graph to a synthesizability score via iterative graph convolution and pooling operations.

The performance of this model is illustrated in **Figure 2a**, where it achieves an area under the curve (AUC) of 0.84 together with an accuracy of 78.0% and a F1 score of 78.8%. These metrics confirm that our optimized ensemble exhibits high sensitivity to potential candidates while effectively ruling out non-viable structures. Building upon these metrics, we perform a large-scale screening over the full candidate set, with the CL score distribution depicted in **Figure 2b**. By applying a stringent threshold of CL score ≥ 0.9, we identify 739 candidates, which serve as a focused pool with high synthesizability potential for subsequent multiferroic analysis.

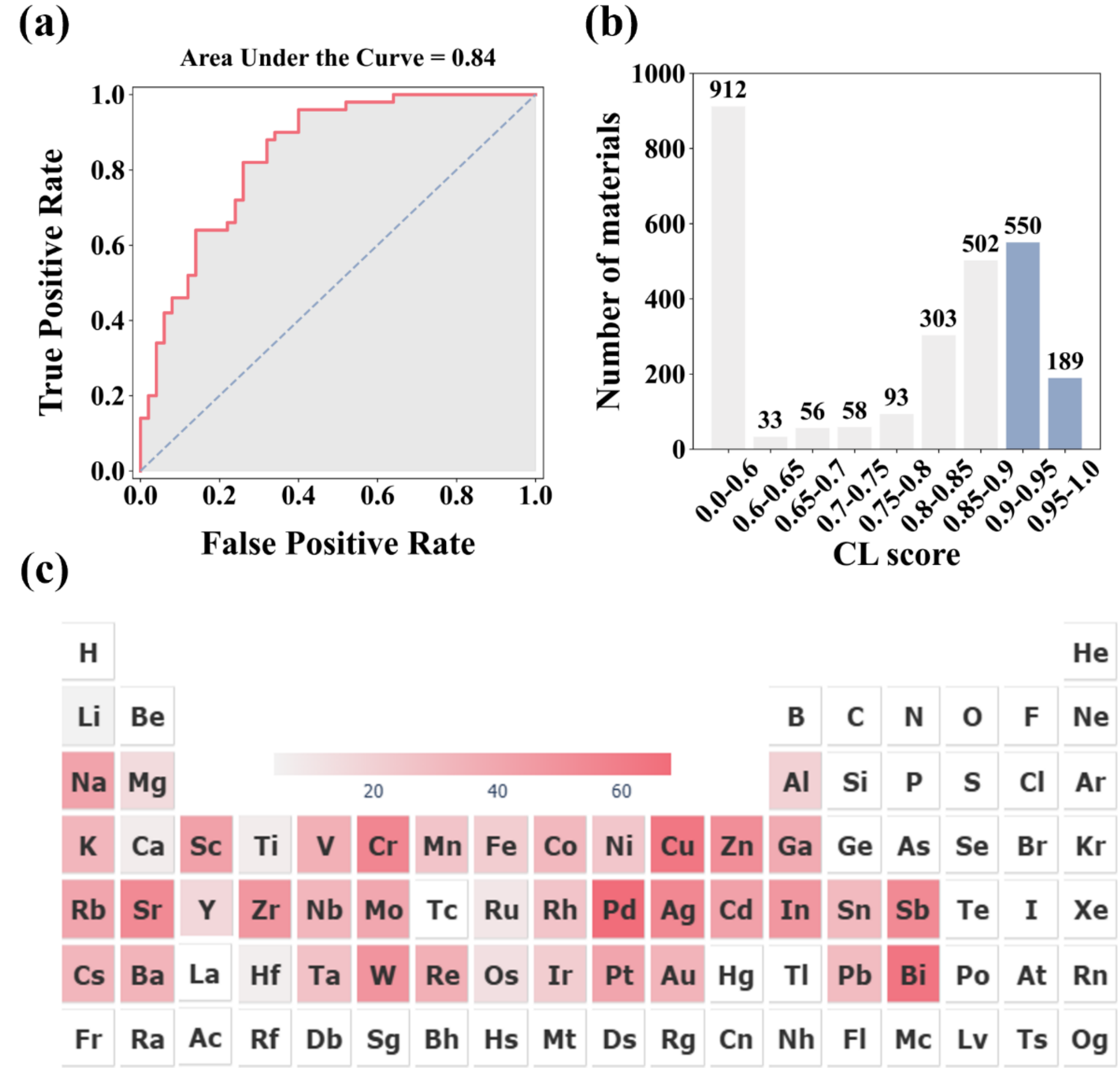

**Figure 2.** (a) Evaluation of the ensemble model performance. (b) CL score distribution for the enumerated structures. (c) Periodic table heatmap showing the elemental abundance in the high-confidence pool.

To elucidate chemical trends within this high-confidence pool, we analyze the occurrence of A/B-site elements and map their distribution onto the periodic table. As summarized in **Figure 2c**, the resulting distribution is highly non-uniform, with frequent appearances of Pd, Bi, and Cu, followed by Cr, Ag, and Sb. Notably, enriched species include $d^{10}$ cations (e.g. Cu, Ag, and Au) and stereochemically active $ns^2$ lone-pair cations (e.g. Bi and Sb), consistent with known stabilization trends in layered chalcogenides and polar distortions.[39-42] Meanwhile, a substantial presence of magnetic transition metals (e.g., Cr and V) indicates that incorporating magnetism remains compatible with the synthesizability filter, enabling the search for intrinsically multiferroic monolayers.

Building upon the 739 high-confidence candidates, we further explore their FE properties. We benchmark multiple regression models and find that gradient boosting (GB) regressor achieves the highest predictive accuracy. The detailed results are shown in **Figure S2**. Accordingly, we employ the GB model with the features listed in **Table S4** to screen candidates with large out-of-plane polarization. To elucidate the origin of the model predictions, we perform Shapley additive explanations (SHAP) analysis to quantify feature contributions.[43] The results presented in **Figure 3a** and 3**b** highlight that the mean atomic radius (abbreviated as Mean AtRad) dominates the model decision, exhibiting an overall negative correlation with polarization. Since this radius is defined as a composition-weighted average, it is naturally governed by the chalcogen site, which accounts for 60% of atoms per formula unit (f.u.). Consistent with this trend, smaller *X*-site anions favor stronger out-of-plane polarization. For example, replacing Se with S increases the polarization from ~1.71 pC/m in $AgBiP_2Se_6$ to ~2.47 pC/m in $AgBiP_2S_6$.[44] In addition, the maximum metal period (denoted as Mag Max Row) shows a significant positive contribution, suggesting that incorporating heavier metals tends to enhance polarization within this family. Together, these attributions provide a practical design guideline for $ABC_2X_6$ candidates, in which large polarization is favored by pairing higher-period metals with smaller chalcogens.

**Figure 3c** summarizes the screened landscape by correlating polarization *P* with switching barrier $E_b$ for the high-throughput-validated 2D ferroelectrics. The identified candidates comprise 45 FE semiconductors, 3 FE metals, and 5 multiferroic semiconductors. A detailed summary of these materials, including CL scores, FE polarization, switching barriers, and bandgaps, is tabulated in **Tables S5-S7**. Several established compounds, including $CuInP_2S_6$,[45-47] $CuInP_2Se_6$,[39] $CuBiP_2Se_6$,[39] and $AgBiP_2Se_6$[40,41] are rediscovered within our high-confidence pool, supporting the reliability of our workflow. Importantly, $AuCrP_2S_6$ distinguishes itself with a substantial $P$ of ~7.46 pC/m and an $E_b$ of ~130 meV. Moreover, **Table S8** and **S9** (supplementary material) demonstrate that $AuCrP_2S_6$ maintains an intrinsic ferromagnetic ground state for both FE and PE phases, highlighting its promise for multifunctional devices.

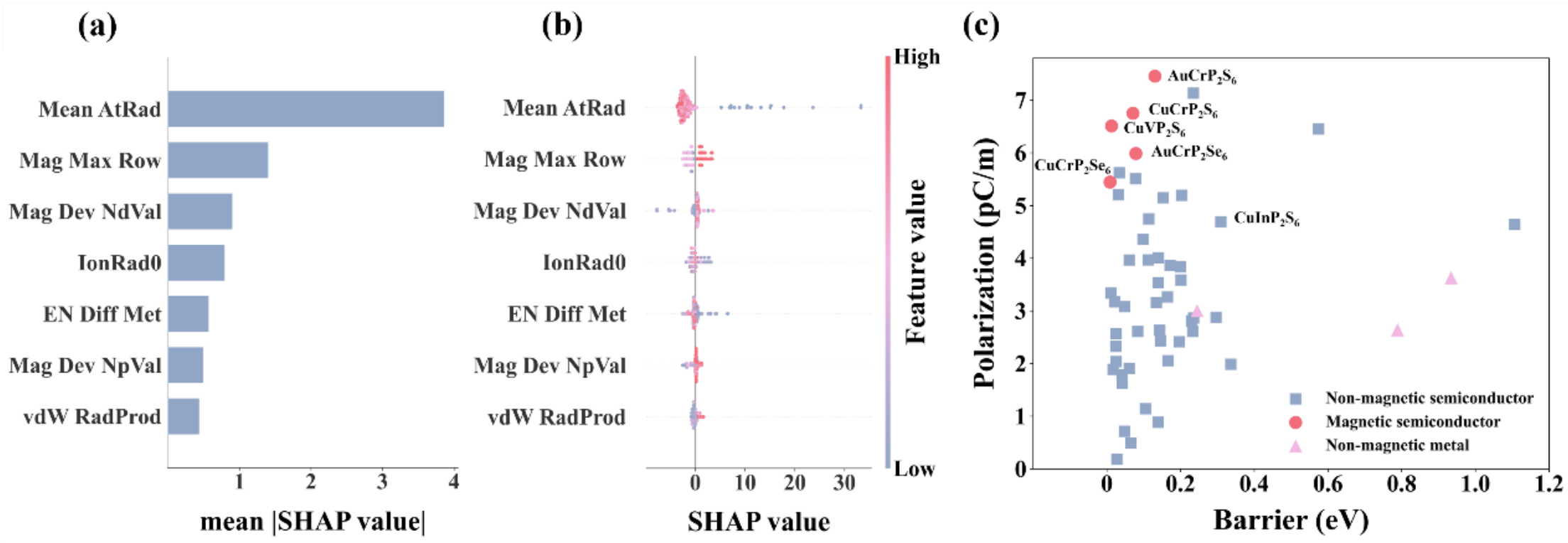

**Figure 3.** (a) Hierarchy of feature importance derived from the mean absolute SHAP value. (b) SHAP summary (beeswarm) plot showing the signed contribution of each feature to the predicted polarization. The horizontal position (SHAP value) indicates whether a feature enhances (positive) or suppresses (negative) the polarization. (c) Screening map of polarization $P$ versus switching barrier $E_b$ for high-throughput validated 2D ferroelectrics.

As depicted in **Figure 4a**, monolayer $AuCrP_2S_6$ adopts a hexagonal lattice in which the $[P_2S_6]^{4-}$ framework encloses the Au and Cr cations. In its FE phase, the Au atom undergoes a large displacement toward one side of the layer, whereas the Cr atom shifts slightly in the opposite direction. This symmetry-breaking distortion produces a sizable out-of-plane polarization and reduces the symmetry to the $C_3$ point group. When the metal cations return to the midplane of the chalcogen framework, the out-of-plane dipole cancels and the polarization vanishes. This configuration corresponds to the PE phase with $D_3$ symmetry, which serves as the non-polar state despite the absence of an inversion center.

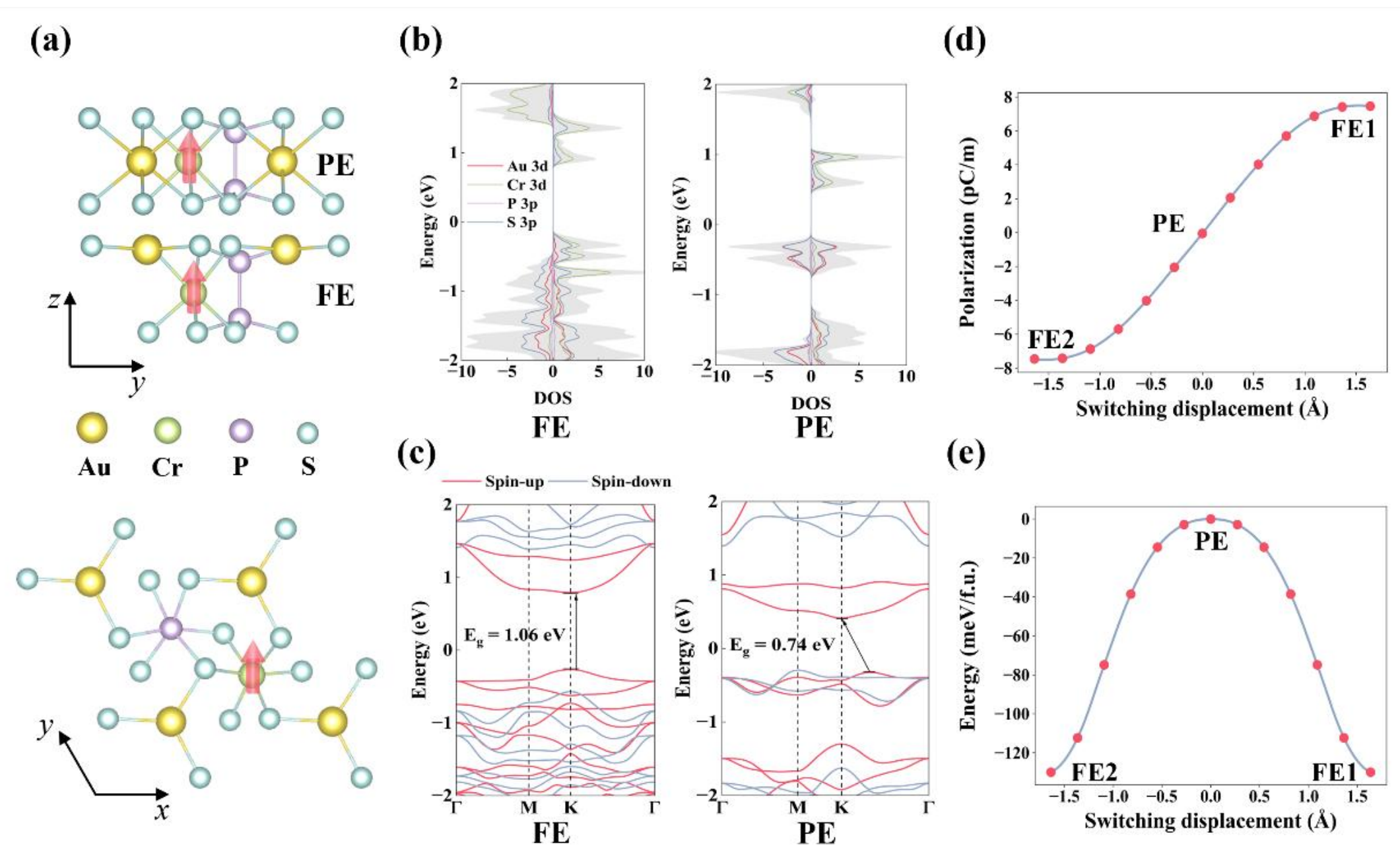

**Figure 4.** (a) Top and side views of the crystal structure, where red arrows denote the spin orientation of Cr atoms. (b) Spin-polarized density of states (DOS) and (c) band structures for the FE and PE phases. (d) Polarization evolution and (e) corresponding double-well energy profile along the minimum-energy switching pathway as a function of the switching displacement.

To further probe its intrinsic electronic properties, we analyze the spin-polarized density of states (DOS) and band structures for both phases, as presented in **Figures 4b** and **4c**. In the FE phase, electronic states near the Fermi level are primarily derived from the hybridization between Cr 3*d* and S 3*p* orbitals. The projected DOS exhibits pronounced exchange splitting, placing the spin-up states closer to the Fermi level than their spin-down counterparts. Accordingly, the fundamental gap is defined by the spin-up channel, with a direct bandgap of 1.06 eV at the *K* point. Upon transitioning to the PE phase, the orbital character of the valence band shifts toward Au 5*d* and S 3*p* states, while the conduction band remains governed by Cr 3*d* and S 3*p* orbitals. Although the phase transition narrows the gap to 0.74 eV and renders it indirect, the exchange splitting associated with the Cr is preserved. This coexistence of a switchable FE polarization and robust exchange-split band edges establish the electronic basis for the coupled optoelectronic response.

To quantify the feasibility of FE switching, we compute the minimum energy pathway using the climbing image nudged elastic band method.[48] **Figures 4d** and **4e** plot the polarization and energy along the switching coordinate, where the switching displacement is defined by the relative out-of-plane separation between the Au and Cr atoms. The polarization evolves smoothly from the negative to the positive FE state through the PE configuration and reaches a spontaneous magnitude of about 7.46 pC/m. This value is comparable to other established 2D ferroelectrics, such as 2D $In_2Se_3$ and $CuInP_2S_6$,[44,48] and significantly exceeds that of sliding ferroelectrics like bilayer $WTe_2$.[49] The energy profile forms a symmetric double well with the PE structure at the saddle point, yielding a switching barrier of approximately 130 meV/f.u. This moderate barrier suggests that the polarization of $AuCrP_2S_6$ is robust against thermal fluctuations yet remains switchable under practical electric fields, making it a promising candidate for nonvolatile memory applications.

Having confirmed the robust multiferroic nature of the $AuCrP_2S_6$ monolayer, we next focus on the shift current to establish a nondestructive readout scheme. As a dominant mechanism of the BPVE under linearly polarized illumination, this nonlinear optical response is intrinsically rooted in the quantum geometry of Bloch electrons.[51-53] The resulting photocurrent is characterized by the optical transition matrix elements and the shift vector, which quantifies the real space displacement of photoexcited carriers during an interband transition. Theoretically, the $C_3$ point group symmetry of the FE phase allows nine independent nonvanishing shift-current components.[54] However, given the 2D geometry and the transverse polarization of the incident light, only two independent components survive, namely $\sigma^{xyy}$ and $\sigma^{yyy}$. Accordingly, we define the transport direction along the $x$-axis and employ $y$-polarized light, thereby identify $\sigma^{xyy}$ as a direct probe of polarization switching. The conductivity tensor is evaluated using maximally localized Wannier functions within the framework of nonlinear optical theory (see **Section S3** in supplementary material for a detailed discussion).

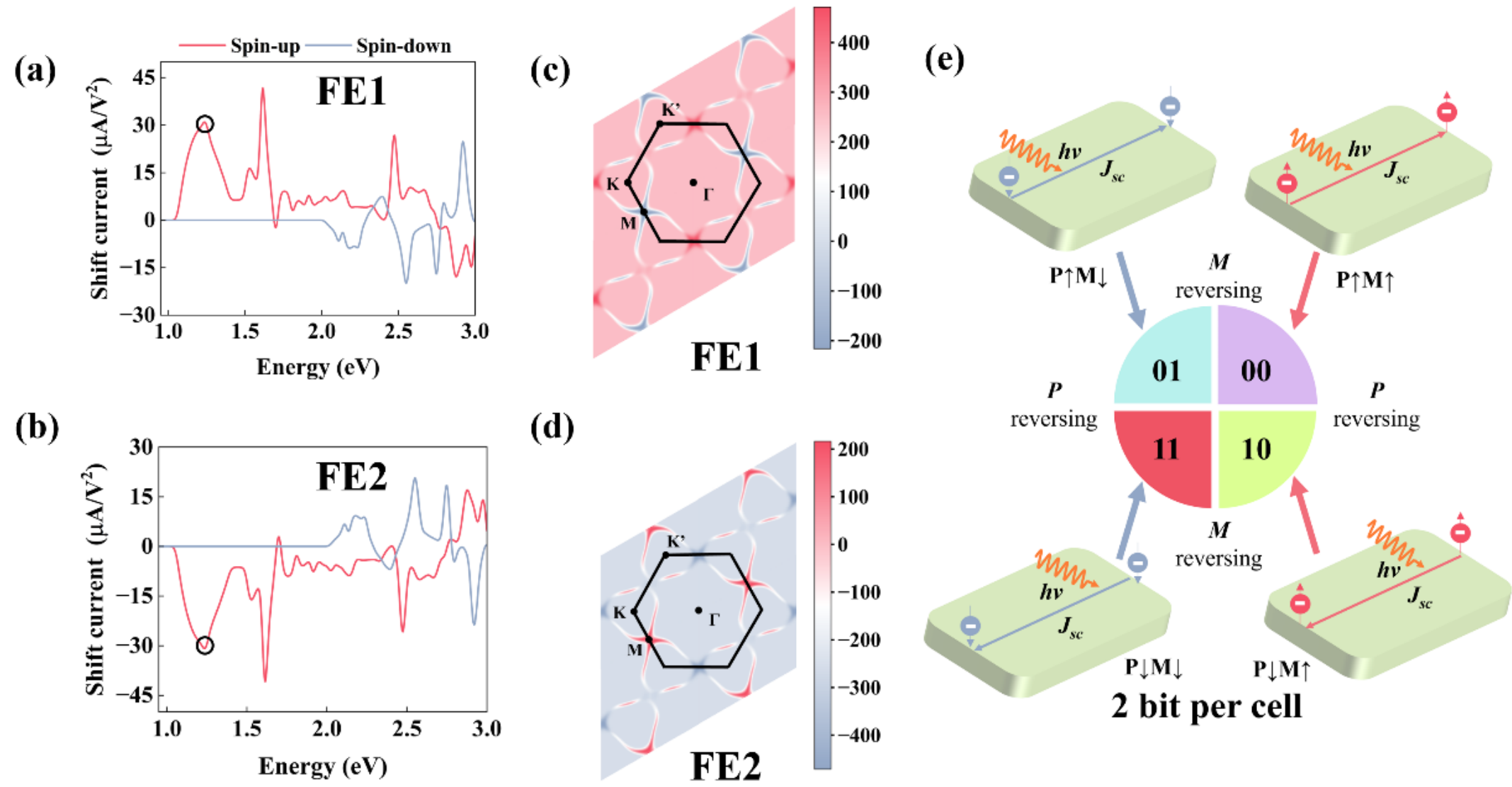

**Figure 5.** Frequency-dependent shift-current spectra for (a) FE1 and (b) FE2 states with the Cr moments initialized in the spin-up configuration. $k$-resolved maps of the shift-current contribution at the peak energy for (c) FE1 and (d) FE2 states. (e) Schematic of a shift-current-enabled four-state nonvolatile memory cell. Switching of $P$ and $M$ enables four distinct logic states (00, 01, 10, 11), corresponding to 2-bit per cell storage.

With the Cr magnetic moments initialized in the spin-up configuration, we compare the frequency-dependent shift-current response for the two opposite polarization states (FE1 and FE2). As shown in **Figures 5a** and **5b**, the spectra are nearly antisymmetric, with equal magnitudes but opposite signs over the 1.0-3.0 eV energy range. This behavior confirms that reversing the polarization $P$ deterministically reverses the photocurrent direction, enabling logic-state encoding. In addition, the spectrum in the 1.0-1.5 eV range is well-defined and dominated by a single pronounced peak near 1.24 eV, which offers a robust and unambiguous window for readout. To trace the microscopic origin of this feature, we examine the momentum-space distribution at this peak energy. As illustrated in **Figures 5c** and **5d**, the $k$-resolved maps for the two FE states exhibit a sign inversion while maintaining the identical magnitude distribution, consistent with the macroscopic reversal of $\sigma^{xyy}$. The analysis further reveals that the optical transitions near the M point provide the primary contribution to the shift current peak.

In addition to polarization sensitivity, the unique spin-resolved band structure of $AuCrP_2S_6$ endows a strong spin selectivity to its shift current. Driven by the exchange splitting, the disparity in bandgaps between the two spin channels ensures that only the channel with the narrower gap supports efficient optical transitions within the 1.0-2.0 eV energy range (see **Figure 5a** and **5b**). Consequently, when the magnetic moments of Cr are initialized in the spin-up configuration, our calculations show that the photocurrent originates entirely from the spin-up channel. Conversely, reversing the magnetic configuration of Cr switches the active contribution to the spin-down channel, as shown in **Figure S5**. Nevertheless, the polarization-induced sign reversal preserves in both cases. This establishes a clean separation of roles in the optical response: the photocurrent direction encodes the polarization state, while the spin character of the carriers is controlled by the magnetic state.

Building upon this mechanism, a four-state nonvolatile memory device is proposed. As illustrated in **Figure 5e**, by assigning the upward orientation of both order parameters ($P$ and $M$) to logic "0" and the downward orientation to logic "1", we establish four distinct multiferroic configurations corresponding to the 00, 01, 10, and 11 states. Electric and magnetic fields can independently switch $P$ and $M$, enabling selective writing of any target logic state. For the readout operation, we exploit the specific symmetry of the photocurrent response. The shift current serves as the probe for the FE bit because inverting $P$ reverses the current direction under fixed illumination. Simultaneously, the magnetic bit is decoded by filtering the carrier transport with spin-selective contacts, where the active spin channel is uniquely locked to the magnetic order. These results demonstrate that monolayer $AuCrP_2S_6$ supports a unified optoelectronic readout in which the combined current direction and spin polarization uniquely identify all four logic states, paving the way toward high-density multistate nonvolatile memory.

In summary, we combine machine-learning screening with first-principles calculations to identify high-confidence two-dimensional vdW multiferroic candidates from the $ABC_2X_6$ family. This strategy enables efficient screening of the vast $ABC_2X_6$ chemical space. Among these candidates, monolayer $AuCrP_2S_6$ stands out as a representative multiferroic semiconductor with a ferromagnetic ground state and switchable out-of-plane polarization. Beyond materials identification, we demonstrate that $AuCrP_2S_6$ enables a unified optoelectronic readout via the shift current. By exploiting the polarization-controlled photocurrent direction and magnetization-selected active spin channel, we propose a device architecture capable of deterministically reading four logic states ($\pm P$, $\pm M$). Our findings not only highlight a practical pathway toward high-density 2-bit memory in the post-Moore era but also establish a generalized paradigm for data-sparse screening, providing a scalable blueprint for the accelerated discovery of functional materials.

See the supplementary material for additional information regarding the PU-Bagging framework, regression-learning framework, first-principles methods, predicted stable FE materials, magnetic ground state determination, and shift current in the spin-down configuration of Cr.

**Acknowledgements**

The authors are grateful for the Shenzhen Natural Science Fund (the Stable Support Plan Program 20231121110218001)

**AUTHOR DECLARATIONS**

**Conflict of Interest**

The authors have no conflicts to disclose.

**DATA AVAILABILITY**

The data that support the findings of this study are available from the corresponding author upon reasonable request.